\documentclass[pra,showpacs,twocolumn]{revtex4}%
\input{psfig.sty}
\usepackage{amsmath}
\usepackage{graphicx}
\usepackage{amsfonts}
\usepackage{amssymb}%

\begin{document}
\title{Non-symmetric entanglement of atomic ensembles}
\date{\today }
\author{A. Kuzmich and T.A.B. Kennedy}
\affiliation{School of Physics, Georgia Institute of Technology,
Atlanta, Georgia 30332-0430}

\begin{abstract}
The entanglement of multi-atom quantum states is considered. In
order to cancel noise due to inhomogeneous light atom coupling,
the concept of matched multi-atom observables is proposed. As a
means to eliminate an important form of decoherence this idea
should be of broad relevance for quantum information processing
with atomic ensembles. The general approach is illustrated on the
example of rotation angle measurement, and it is shown that the
multi-atom states that were thought to be only weakly entangled
can exhibit near-maximum entanglement.
\end{abstract}
\maketitle

Progress in quantum communication and information processing
depends on practical schemes for the massive entanglement of
quantum systems. Advances based on the entanglement of atomic
ensembles are undergoing rapid development, and involve
independent approaches toward the entanglement of continuous
light-atom variables (see e.g., Ref. \cite{kluwer} and references
therein), and discrete variables \cite{kuz-nature,zoller}. Recent
progress suggests that in the short term major advances in quantum
information processing with atomic ensembles are likely.

What makes quantum information processing with atomic ensembles
very attractive is the relative simplicity of the experimental
interface. As opposed to the case of single atoms interacting with
single photons, the use of atomic ensembles does not require the
technology of strong coupling cavity QED, and a free space
interaction between light and atomic ensemble is sufficient for
massive entanglement. In the latter case the conditions for a
sufficiently strong interaction reduce to a high on-resonance
optical density of the atomic sample, a condition which is
relatively easy to achieve experimentally. Therefore, the field of
atomic ensemble entanglement shows enormous promise as a practical
way to implement quantum information processing.

On the other hand, a well-known difficulty of the field has been
how to realize the required symmetric character of the atom-photon
interaction in the presence of inhomogeneities in the local
atom-field coupling. Inhomogeneity of this kind is a potentially
fatal problem for applications of atomic ensembles to quantum
information processing [4].

This apparent problem is rooted in a widely accepted notion that
the multi-atom states have to exhibit a high degree of symmetry in
order to be strongly entangled. Implicit in these arguments has
been the fact that conventional entanglement measures, such as the
spin squeezing parameter, are themselves symmetric
\cite{yurke,kitagawa,wineland92,wineland94,asorensen}.  To the
best of our knowledge, in all of the atomic ensemble entanglement
work such symmetric entanglement measures have been used
\cite{grom,kuz-EPL,
kuz-PRL,klaus-a,hald,kuz-PRA,kmb,fl,atomtele,luming,julsgaard,stewart,bouchoule,
dilisi,qse,kuz-nature,lukin-science,klaus1,klaus}. The symmetry of
entanglement measures is directly related to the fact that
additive $N$ atom observables of the form $ \hat A =
\sum_{k=1}^{N} \hat a_k$, which are symmetric with respect to
permutation $ \hat a_k \leftrightarrow \hat a_l$ of any two single
particle operators satisfying identical commutations relations,
were thought to provide a proper connection to the experimental
situation. Even for modest degrees of asymmetry of multi-atom
quantum states, the fact that $N >> 1$ imposes severe limitations
on the degree of entanglement of such symmetric observables
\cite{liyou}.

There is, however, no a priori reason that the multiatom
observables, and consequently, entanglement measures, should
exhibit this kind of symmetry.
We would like to argue in favor of a more operational approach, in
which the measurement procedure itself essentially defines the
entanglement measure. In atomic ensemble entanglment experiments
the multi-atom observables are typically non-symmetric, because
the atoms are distributed in space and couple to the local value
of the electric field mode
\cite{hald,kmb,julsgaard,kuz-nature,lukin-science}. Instead of
symmetric operators like $\hat A$, a related non-symmetric
observable $\tilde{A} = \sum_{k=1}^{N} g_k \hat a_k$ should be
considered, where the $g_k$ are a set of numbers whose deviation
from unity is associated with the inhomogeneity. We will show that
there is no asymmetry induced decoherence in the entanglement of
such matched observables.

Our conjecture is that for any ensembles-based quantum information
protocol, the asymmetry of the atomic ensemble cancels out and
near-maximum entanglement is observed, provided every step of the
ensemble-based interaction {\it is matched} to each other. In a
sense the classical noise associated with the observable
asymmetry, or equivalently the atom-field coupling inhomogeneity,
can be explicitly accounted for and subtracted out.

While the notion of matched observables is quite general, for the
sake of clarity we will illustrate it on a particular atomic
ensemble scheme: the sensitivity of rotation angle measurement.
First let us recall the general ideas involved. The simplest kind
of rotation measurement involves preparing an input quantum state
of an $N$ particle system, passing it through an interferometer
where a phase rotation is encoded, and then measuring an output
observable sensitive to the phase rotation. The action of the
quantum interferometer (photon or massive particle) can quite
generally be understood in terms of the rotation of a pseudo
angular momentum operator ${\bf \hat F }$, comprised of an
ensemble of N spin 1/2 systems ${\bf \hat F }= \sum_{k=1}^{N} {\bf
\hat f^k}$ \cite{YMcCK86}. The sensitivity of the rotation
measurement is evidently conditioned on the quantum fluctuations
of the input state of ${\bf \hat F }$. For a coherent spin state
(CSS), a non-entangled state, the rotation sensitivity is given by
the so-called standard quantum limit $\delta \phi \sim 1/
\sqrt{N}$. If instead we prepare a spin squeezed state (SSS), a
massively entangled state of $N$ particles, the phase sensitivity
improves and can approach the Heisenberg limit  $\delta \phi \sim
1/ N$ \cite{YMcCK86}. This example rather clearly illustrates the
practical consequences of entanglement in quantum measurements.
The crucial issue, then, is how to prepare the entangled SSS? One
proposal involves allowing the spin ensemble to interact for a
certain time with an auxiliary ensemble of n spin 1/2 particles
${\bf \hat s^i}, i=1,...,n$, prior to entering the interferometer
\cite{kuz-EPL}. Two essential new elements of this scheme, in
addition to the action of the interferometer itself, should be
noted. The interaction between the two spin ensembles is designed
to prepare a SSS, and for this purpose a quantum non-demolition (
QND ) interaction is ideal. Moreover, an external measurement on
the auxiliary spin is required ${\bf \hat S} = \sum_{i=1}^{n} {\bf
\hat s^i}$, in order to condition the input state to the
interferometer. We shall also see later that entanglement of the
atomic spin ensemble by means of a QND interaction with an
ancilliary ensemble is not essential. One can also exploit
entanglement between the two ensembles when the atomic ensemble is
in a separable state. However, we will first concentrate on the
former case.

The discussion hides an important practical problem, however, the
difficulty of implementing the desired QND interaction. It is at
this point that the issue of non-symmetric observables arises in
practice. Analysis of the example of rotation angle measurement
demonstrates how random noise associated with the asymmetry feeds
into the measurement preventing a sensitivity approaching the
Heisenberg limit. Consider two ensembles of spin 1/2 systems, the
{\it atoms} ${\bf \hat f^i}, i=1,...N$ which are passed through
the interferometer, and the {\it photons} ${\bf \hat s^i},
i=1,...n$ which act as the auxiliary state preparation device.
While at this stage we use the names atoms and photons merely as
labels, we have in mind practical quantum optical systems where
the angular momenta do correspond to a collection of atoms and a
single mode probe light field, respectively. The total angular
momenta are represented by ${\bf \hat S} = \sum_{i=1}^{n} {\bf
\hat s^i}$ and ${\bf \hat F }= \sum_{k=1}^{N} {\bf \hat f^k}$. We
assume that initially the states of both of the ensembles are
uncorrelated CSS, in which the average values of $\bf \hat S$ and
$\bf \hat F$ are directed along the x-axis.

The interaction between the atoms and photons is designed in order
to entangle the ensemble of atomic spins, and is given by the QND
interaction
\begin{equation}  \label{QNDHamil}
\hat H= \hbar \Omega \sum_{i=1,k=1}^{n,N} g_k\hat s^i_z \hat f^k_z
= \hbar \Omega \hat S_z \tilde F_z
\end{equation}
where $\Omega $ is a frequency, the $g_k$ are dimensionless
coupling weights, and $\tilde F_z = \sum_{k=1}^{N} g_k \hat f_z^k$
is a non-symmetric atomic operator unless all of the $g_k$ are
equal. In a realistic experimental scenario coupling weights of
varying magnitude arise when an interaction of this kind is
created using an off-resonant interaction between a single mode
light field and a collection of atoms. Then, the distribution of
coupling weights $g_k$ maps out the variation of mode intensity
seen by the spatially distributed atoms $k=1,... N$.

If the atom and photon spin ensembles interact for a time $\tau$
under Eq.~(\ref{QNDHamil}), the atomic spins evolve according to
$\hat f^k_x(\tau) = \cos(g_k\chi \hat S_z)\hat f^k_x - \sin
(g_k\chi \hat S_z) \hat f^k_y$, $\hat f^k_y(\tau) = \sin (g_k\chi
\hat S_z) \hat f^k_x + \cos (g_k\chi \hat S_z)\hat f^k_y$, $\hat
f^k_z(\tau) = \hat f^k_z$, where $\chi = \Omega \tau$. The
individual spins have a dispersion of frequencies associated with
the distribution of weights, while the collective spin ${\bf \hat
S}$ satisfies
\begin{equation}  \label{jout}
\left(
\begin{array}{c}
\hat S_{x~} \\
\hat S_{y~} \\
\hat S_{z~}
\end{array}
\right)({\tau }) = \left(
\begin{array}{ccc}
\cos(\chi {\tilde{F_{z}}}) & - \sin(\chi {\tilde{F_{z}}}) & 0 \\
\sin(\chi {\tilde{F_{z}}}) & \cos(\chi {\tilde{F_{z}}}) & 0 \\
0 & 0 & 1
\end{array}
\right)\left(
\begin{array}{c}
\hat S_{x~} \\
\hat S_{y~} \\
\hat S_{z~}
\end{array}
\right).
\end{equation}
The atoms are then passed into the interferometer where a phase
rotation $\phi$ is imposed by means of a rotation about the
y-axis, so that $\hat f^k_{a,out} = e^{i\phi \hat F_y(\tau)} \hat
f^k_{a}(\tau) e^{-i\phi \hat F_y(\tau)}$, $a=x,y,z$. To determine
the phase $\phi$ we could measure the z component of total angular
momentum at the output $\hat F_{z,out} = \hat F_z \cos \phi - \sin
\phi \sum_{k=1}^{N} \left( \cos (g_k\chi \hat S_z)\hat f^k_x -
\sin (g_k\chi \hat S_z) \hat f^k_y \right)$. In fact we should
really like to measure $\hat F_{z,out}- \hat F_{z}(\tau) = \hat
F_{z,out}- \hat F_{z}$ the change in angular momentum due to the
interferometer, cancelling any superfluous noise carried in the
SSS input. The QND interaction enables us to obtain information
about $\hat F_{z}$, since $\hat S_y(\tau) \approx \chi \tilde F_z
\hat S_x+ \hat S_y$, for $\chi \triangle F_z = \chi \sqrt{N / 4}
<< 1$. Hence, using $\langle \hat S_x \rangle_{CSS} = N/2$, it
follows that the operator $2 \hat S_y(\tau) / (n \chi) \approx
\tilde F_z (2 \hat S_x / n) + 2 \hat S_y / (n \chi)$, is linear in
$\tilde F_z$ with a coefficient of unity ( the coherent state is
an eigenstate of $\hat S_x$, and thus it can be replaced by the
c-number $n/2$). This operator is as close to $\hat F_z$ as we can
get by measuring a component of the auxiliary ensemble. Note that
the observable which appears is the non-symmetric partner $\tilde
F_z$ rather than the ( symmetric ) angular momentum $\hat F_z$,
and thus classical noise associated with the distribution of
weights $g_k$ feeds into measurements of the observable $\hat
F_z^{\prime} \equiv \hat F_{z,out} - 2 \hat S_y(\tau) / (n \chi)$.

To determine the sensitivity of rotation measurements we now
evaluate
\begin{equation}
\delta \phi = \left( \sqrt{\langle (\Delta \hat F_z^{\prime})^2
\rangle } / \big|{\frac{d \langle \hat F_z^{\prime}
\rangle}{d\phi}}\big| \right)_{\phi = 0} .
\end{equation}
It is sufficient to evaluate  $\langle \hat F_z^{\prime} \rangle$
correct to first order in $\phi$, i.e., $\langle \hat F_z' \rangle
\approx -(N \phi/2) \langle e^{-\xi g^2/2} \rangle
  = -(N \phi/2) \frac{ e^{-\xi(1 + \Delta g^2 \xi)^{-1}/2} }{\sqrt{1 +
\Delta g^2 \xi} }$ for $g_k^2 \chi^2 << 1$ and $ \xi (\Delta g)^2
<< 1 $, where $\xi \equiv n \chi^2 /4$. Here we have used
properties of the initial coherent spin state, and assumed that
the weights $g_k$ are gaussian distributed with mean value
$\langle g \rangle = 1$, and variance $(\Delta g)^2 = \langle
g^2\rangle -1$.

The variance $\langle (\Delta \hat F_z^{\prime})^2 \rangle $ may
be computed at $\phi = 0$, and using conservation of $\hat F_z$ it
reduces to $\langle (\Delta \hat F_z^{\prime})^2 \rangle = \langle
(\Delta \hat F_z)^2 \rangle + 4 \langle \hat \hat
S_y^2(\tau)\rangle/(n^2 \chi^2) - 2 \left( \langle \hat F_z \hat
S_y(\tau) \rangle + c.c \right)/ (n \chi)$. The first and second
members have shot noise contributions which are cancelled by the
third term, as designed, leaving residual quantum fluctuations.
Explicitly $ 4 \langle \hat S_y^2(\tau)\rangle/(n^2 \chi^2)
\approx \left( N + 1/\xi + N (\Delta g)^2 \right) / 4$, assuming
$N \chi^2 << 1$, where the three terms represent shot noise,
quantum fluctuations associated with the input entangled state and
classical noise due to the distribution of weights, respectively.
Hence, cancelling the atomic shot noise, we find $(\Delta \hat
F_z^{\prime})^2 = \left(1 + N \xi (\Delta g)^2 \right)/ (4 \xi)$,
which gives a phase error of
\begin{equation}
\delta \phi(\xi) = \sqrt{\left[ 1 + N \xi (\Delta g)^2
\right]\left[ 1 + \xi (\Delta g)^2 \right]} \frac{e^{\xi \left[
1+\xi (\Delta g)^2\right]^{-1} /2}}{N \sqrt{\xi}}.
\end{equation}
In the symmetric limit $\Delta g = 0$, $\langle g^2 \rangle = 1$,
this reduces to $ \delta \phi(\xi) =  e^{\xi/2} / \left(N
\sqrt{\xi}\right) $. It takes a minimum value approaching the
Heisenberg limit for $\xi = 1$, where $\delta \phi_{min} =
\sqrt{e}/N$ \cite{kuz-EPL}. Note that $\xi = 1$ implies $ n \chi^2
= 4$, while we have also assumed $N \chi^2 n^{1/2} << 4$ to obtain
the approximate expression for the variance. The two conditions
are consistent provided the number of photons greatly exceeds the
number of atoms $\sqrt{n} >> N$. In order to approach the
Heisenberg limit, therefore, it is clear that we must satisfy
$(\Delta g)^2 << \frac{1 }{N}$
 a criterion which is very difficult to achieve for a macroscopic sample of
  atoms where typically $10^6 < N < 10^9$.

 The procedure we have described is based on
 a measurement of $\hat F_z$, but the QND interaction does not correlate the
 auxiliary spin angular momentum $\hat S_z$ directly to  $\hat F_z$,
 rather to its non-symmetric partner
 $\tilde F_z$. Hence the inhomogeneity noise will dominate the phase
 measurement error unless very strict limits on the distribution
 variance are satisfied. We have explicitly demonstrated the
 conventional wisdom that asymmetry washes out the entanglement.
  The important question is, can the
 entanglement be distilled out if suitable multi-particle
 measurements are made to cancel the classical noise?
 Our answer is in the affirmative.
 The crucial point is that since the form
 of the QND interaction correlates the photon spin to $\tilde
 F_z$, we require a multi-particle measurement which gives $\tilde
 F_z$ also at the output of the interferometer. The difference
 signal can then in principle cancel the coupling distribution noise. This strategy is quite
 clear from the operational point of view taken here, but runs counter to
 standard theoretical measures of entanglement based on symmetric operators, such as
 the spin squeezing parameter $\eta = \sqrt{N} \delta \phi$, which
 are implicitly sensitive to distribution noise.

 To determine which multi-particle measurements will achieve our
 goal, we consider the operator $\tilde F_z$ at the interferometer
 output. To first order in $\phi$
 \begin{equation}
\tilde F_{z,out} = \tilde F_z - \phi \tilde F_x(\tau)
 \end{equation}
Now we introduce a second set of auxiliary spins ${\bf \hat j^i},
i = 1,..., n$ which are coupled with the atomic spin operators at
the interferometer output in exactly the same way as spin ${\bf
\hat S}$, i.e., $H_{int} = \hbar \Omega \hat J_z \tilde F_z$,
where ${\bf \hat J} = \sum_{i=1}^{n} {\bf \hat j^i}$. In practice
this ensemble represents a second pulse of light in the same
spatial mode of the electromagnetic field, and we assume that it
is prepared in a CSS pointing in the positive x direction, so that
$\langle \hat J_x \rangle = n /2$. Thus after interacting for a
time $\tau$ with the atoms we have to sufficient accuracy for $N
\chi^2 << 1$ that $\hat S_y(\tau) = \hat S_y + \chi \tilde F_z
\hat S_x$, and $\hat J_y(\tau) = \hat J_y + \chi \tilde F_{z,out}
\hat J_x = \hat J_y + \chi \tilde F_z \hat J_x - \phi \tilde
F_x(\tau)\hat J_x$.

Let us now consider the difference signal $\hat A(\phi) \equiv
\hat J_y(\tau) - \hat S_y(\tau)$, so that
\begin{equation}
\hat A(\phi) = \hat J_y - \hat S_y + \chi \tilde F_z \left( \hat
J_x - \hat S_x\right) - \phi \chi \tilde F_x(\tau) \hat J_x.
\end{equation}
As the auxiliary spins are prepared in eigenstates of $\hat S_x$
and $\hat J_x$, respectively, the variance in $\hat A(\phi=0)$
reduces to the sum of shot noise contributions from the two
auxiliary spin systems $ \langle \left( \Delta \hat A(\phi=0)
\right)^2 \rangle = \langle (\Delta \hat J_y)^2\rangle + \langle
(\Delta \hat S_y)^2 \rangle = n/2, $ and the weight distribution
noise has cancelled out. The ensemble average of the differential
change in signal at $\phi = 0$ measures $\tilde F_x(\tau)$
directly, i.e.,
\begin{eqnarray} \nonumber
\left( \frac{d}{d\phi} \langle \hat A(\phi) \rangle \right)_{\phi
= 0} &&= - \chi \langle \hat J_x \rangle \langle \tilde F_x(\tau) \rangle\\
&&= - \frac{\chi n N}{4} \frac{ e^{-\xi [1+\xi \left( \Delta g
\right)^2]^{-1} /2} }{\left[1 + \xi \left( \Delta g \right)^2
\right]^{3/2}},
\end{eqnarray}
and the corresponding phase error is given by
\begin{equation}
\delta \phi(\xi) = \sqrt{\frac{2}{\xi}} \ \frac{ \left( 1+\xi
\left( \Delta g \right)^2 \right)^{3/2}}{N } \ e^{-\xi [1+\xi
\left( \Delta g \right)^2]^{-1} /2}.
\end{equation}
For $(\Delta g)^2 << 1$, the minimum phase error approaches the
Heisenberg limit $\delta \phi_{min} = \sqrt{2e}/N$, where the
factor of $\sqrt 2$ arises from the independent shot noise
contributions of the two auxiliary spin ensembles. It is
interesting, to note that even when $\Delta g \rightarrow
\sqrt{\langle g^2 \rangle}$, the phase error still scales as
$1/N$. In other words this measurement scheme could operate close
to the Heisenberg limit of phase measurement accuracy under the
same conditions where a symmetric squeezing parameter predicts
little or no entanglement. The measurements are appropriately
\textit{matched} to the non-symmetric observable of interest, and
the corresponding phase error yields an improved, operational
measure of entanglement.

It is possible to avoid using the 2nd photon ensemble ${\bf \hat
j^i}, i = 1,..., n$, altogether. Instead,  the first photon
ensemble ${\bf \hat s^i}, i = 1,..., n$ can be stored after the
first QND interaction. After the phase rotation $\phi$, the photon
ensemble is coupled to the atomic spins  with interaction strength
$\chi _1 =-\chi $ (this can be effectively achieved by flipping
the z-component of either atom or photon spin before the
interaction takes place). The $\hat S_y ^f$ component of the
photon ensemble serves as the $\phi $-dependent observable $\hat
B(\phi )\equiv \hat S_y ^f$. We find $ \hat B(\phi ) = \hat S_y +
\left( \chi \tilde F_z -\chi \cos (\phi ) \tilde F_z +\chi \sin
(\phi ) \tilde F_x \right)\hat S_x. $ It follows that $\langle
\hat B(\phi ) \rangle = - \langle
 \hat A(\phi)  \rangle$, and
 $\langle \left( \Delta \hat B(\phi=0) \right)^2
\rangle =\frac{1}{2}\langle \left( \Delta \hat A(\phi=0) \right)^2
\rangle $, since now the shot noise of only one photon ensemble
contributes. The phase accuracy  $\delta \phi (\xi )$ is improved
by a factor of $\sqrt{2}$ over that given by Eq.(8). It is
straightforward to check that in this situation spin squeezing
parameters \cite{kitagawa,wineland94,zoller} for the atomic
ensemble always exceed or equal unity at all times even for
perfectly symmetric coupling. Thus the atomic state is not
``spin-squeezed", or entangled. Nevertheless, Heisenberg-limited
performance of the atomic interferometer is achieved. Therefore,
we argue that atom-photon entanglement in itself is the primary
reason behind the improved accuracy of the considered QND-enhanced
atomic interference measurement; the measurement-induced atomic
squeezing is secondary and its use may be  avoided.

Our results are of importance for the synergy of the fields of
cavity QED \cite{kimble} and of atomic ensembles \cite{kluwer}.
Using a cavity of finesse $F$ would effectively increase the
strength of the ensemble-light interaction by the factor of $F$
\cite{kuz-EPL}. It has been suggested in the context of atomic
ensembles that a ring cavity with running wave cavity modes and
large Gaussian-waists should be used, so that all atoms see the
same electric dipole mode coupling intensity
\cite{kuz-EPL,klaus1,klaus}. On the contrary, here we put forward
an argument that suggests even standing-wave cavity modes can be
used to realize a QND atom-light interaction.  In this case
$\langle (\Delta g)^2\rangle
 \rightarrow \langle g\rangle ^2$, but as we have shown, it is
 still possible to approach the
Heisenberg limit. In present-day experiments various kinds of
decoherence would practically limit the amount of entanglement
before the Heisenberg limit is reached. With that in mind, we
could say that standing wave probe-fields are just as good as
running-wave ones. The existing high finesse cavities typically
used in cavity QED are all of standing wave type, and the cavity
modes have a sinusoidal spatial dependence along the cavity axis
\cite{orozco,chapman,mckeever,pinkse}.

To what extent is the matching of observables feasible in
practice? Let us consider the most demanding case of the standing
wave cavity interacting with an ensemble of cold atoms. The atomic
velocity would be on the order of a few cm/s, thus if the time
interval between the light pulses $<$ 10 $\mu $s, the two
measurements would be very well matched. For comparison, in recent
experiments that demonstrated atomic memory effects in atomic
ensembles \cite{kuz-nature,lukin-science} the relevant time
intervals were $<$ 1$\mu$s. In the case of a running-wave probe
field, where the typical length of intensity variations would be a
fraction of a centimeter, time intervals as long as 0.1 s would be
possible.

 Our results should have broad relevance for situations involving
 multiatom entangled states. By presenting a strategy to eliminate
 the noise due to asymmetry of atomic states, we have significantly
 simplified the path towards use of
atomic ensembles for quantum information processing. Quantum
metrology applications such as atomic clocks, gravimeters, and
atomic electric dipole moment searches would also benefit from
enhancement in precision that entangled atomic states can provide.
Beyond the field of atomic ensembles, we hope that this work opens
a discussion of multi-particle entanglement outside the symmetric
entangled states that have been considered so far.

A.K. gratefully acknowledges illuminating discussions with D.
Matsukevich. T.A.B.K. acknowledges support from MUIR and the
hospitality of the University of Insubria where some of the work
was carried out. This research was supported by NASA.


\begin{thebibliography}{99}
\bibitem{kluwer} A. Kuzmich and E. S. Polzik, in {\it Quantum information with continuous
variables}, eds S. L. Braunstein and A. K. Pati, Kluwer, 2003.
\bibitem{kuz-nature} A. Kuzmich et al., Nature (London), {\bf 423}, 731 (2003).
\bibitem{zoller}  L.-M. Duan et al.,
Nature, {\bf 414}, 413 (2001).
\bibitem{liyou} C. P. Sun et al., Phys. Rev. A 67, 063815 (2003).
\bibitem{yurke} B. Yurke, Phys. Rev. Lett. {\bf 56},
1515 (1986).
\bibitem{kitagawa} M. Kitagawa and M. Ueda, Phys. Rev. Lett.
{\bf 67}, 1852 (1991); Phys. Rev. A {\bf 47}, 5138 (1993).
\bibitem{wineland92} D. J. Wineland et al., Phys. Rev. A {\bf 46}, R6797 (1992);
\bibitem{wineland94} D. J. Wineland at al. Phys. Rev. A {\bf 50}, 67 (1994).
\bibitem{asorensen} A. S\o rensen et al., Nature {\bf 409}, 63 (2001).
\bibitem{grom} G.F. Grom and A. M. Kuzmich, JETP Lett. {\bf 61}, 900
(1995).
\bibitem{kuz-PRL} A. Kuzmich et al., Phys. Rev. Lett. {\bf 79}, 4782 (1997).
\bibitem{kuz-EPL} A. Kuzmich et al. Europhys. Lett. {\bf 42}, 481 (1998).
\bibitem{klaus-a} K. M\o lmer, Eur. Phys. J. D {\bf 5}, 301 (1999).
\bibitem{hald}  J. Hald et al.,
Phys. Rev. Lett. {\bf 83}, 1319 (1999).
\bibitem{kuz-PRA}   A. Kuzmich et al., Phys. Rev. A {\bf 60}, 2346 (1999).
\bibitem{kmb}  A. Kuzmich et al., Phys. Rev. Lett. {\bf 85}, 1594 (2000).
\bibitem{fl}  M. Fleischhauer and M.D. Lukin,  Phys. Rev. Lett. {\bf 84}, 5094 (2000).

\bibitem {atomtele} A. Kuzmich and E. S. Polzik, Phys. Rev. Lett.
{\bf 85}, 5639 (2000).
\bibitem {luming} L.-M. Duan et al., Phys. Rev. Lett.
{\bf 85}, 5643 (2000).

\bibitem{julsgaard} B. Julsgaard et al., Nature {\bf 413}, 400 (2001).
\bibitem{stewart} S.D. Jenkins and T.A.B. Kennedy, Phys. Rev. A
{\bf 66}, 043621 (2002).
\bibitem{bouchoule}I. Bouchoule and K. M\o lmer, Phys. Rev. A {\bf 65}, 041803
(2002).
\bibitem{dilisi}A. Di Lisi and K. M\o lmer, Phys. Rev. A 66,
052303 (2002).

\bibitem{qse}  L.-M. Duan,  Phys. Rev. Lett., {\bf 88},
170402 (2002).


\bibitem{lukin-science}C. H. van der Wal et al., Science, {\bf 301}, 196 (2003).

\bibitem{klaus1}A. S. S\o rensen and K. M\o lmer, Phys. Rev. A {\bf 66}, 022314
(2002).
\bibitem{klaus} A. S. S\o rensen and K. M\o lmer, Phys. Rev. Lett. {\bf 90}, 127903
(2003).


\bibitem{YMcCK86}  B. Yurke et al., Phys.
Rev. A {\bf 33}, 4033 (1986).

\bibitem{kimble} H.J. Kimble, Phys. Scr. {\bf 76}, 127 (1998).

\bibitem{orozco} W. P. Smith et al., Phys. Rev. Lett. {\bf 89}, 133601 (2002).
\bibitem{chapman} J.A. Sauer et al.,  Los Alamos
preprint quant-ph/0309052.
\bibitem{mckeever} J. McKeever et al., Phys. Rev. Lett. {\bf 90},
133602 (2003).
\bibitem{pinkse}P.W.H. Pinkse et al., Nature (London) {\bf 404}, 365 (2000).
\end{thebibliography}
\end{document}